\documentclass[preprint,showpacs,prl,twocolumn,10pt]{revtex4}%
\usepackage{amsfonts}
\usepackage{amsmath}
\usepackage{amssymb}
\usepackage{graphicx}%
\begin{document}
\preprint{ }
\title{The Electron-Glass in samples approaching the Mesoscopic regime}
\author{V. Orlyanchik and Z. Ovadyahu}
\affiliation{Racah Institute of Physics, The Hebrew University, Jerusalem 91904, Israel }

\begin{abstract}
We study the dependence of the glassy properties of strongly localized
$In_{2}O_{3-x}$ films on the sample lateral dimensions. Characteristic
mesoscopic effects such as reproducible conductance fluctuations (CF) are
readily observable in gated structures for sample size smaller than 100$~\mu
m$ measured at 4~K, and the relative amplitude of the CF decreases with the
sample volume as does the flicker noise. By contrast, down to sample size of
few microns, the non-equilibrium features that are attributed to the
electron-glass are indistinguishable from those observed in macroscopic
samples, and in particular, the relaxation dynamics is independent of sample
size down to 2$~\mu$m. In addition, The usual features that characterize the
electron-glass including slow-relaxation, memory effects, and full-aging
behavior are all observed in the `mesoscopic' regime, and they appear to be
independent of the conductance fluctuations.

\end{abstract}
\maketitle

\section{Introduction}

Non-ergodic transport properties of Anderson insulators have been reported in
a number of systems studied at low temperatures \cite{1,2,3,4}. When excited
from equilibrium by e.g., a sudden change of a gate voltage, the conductance
of the system increases, and this excess conductance $\Delta G$ persists for
long times (in some cases, days) after the excitation. Vaknin \textit{et al}
\cite{5} argued that such extended relaxation times as well as various memory
effects exhibited by these systems are difficult to explain unless
electron-electron correlations play a decisive role \cite{6}. For a medium
that lacks metallic screening, it is natural to assume that the Coulomb
interaction is relevant. This point of view was indeed taken in some
theoretical models where the long range nature of the Coulomb interaction was
used to justify a mean-field treatment \cite{7}.

In this work we explore the behavior the electron-glass when the system size
approaches the mesoscopic regime of hopping conductance. The study is
motivated by two main reasons. One goal is to assess the spatial range of
correlation in the electron glass by monitoring the dynamics as function of
system size (which seems a viable approach when dealing with a long range
interaction like the un-screened Coulomb potential). This is based on the
notion that the slow relaxation is presumably a result of a many-particle
scenario \cite{6,8}, and therefore the dynamics should become size dependent
below a certain scale \cite{9}. Another reason to study small systems is that
it should shed light on the interplay between the electron slow-dynamics and
the ionic one which may be reflected in the temporal aspect of some specific
mesoscopic effects.

Our main finding is that at least down to sample size $L$ of few microns the
basic electron-glass properties, namely, slow-relaxation, memory, and aging
remain essentially the same as in macroscopic samples. In particular, the
characteristic relaxation time, observable in relaxation experiments of
$\Delta G,$ shows no scale dependence when the sample size is changed from
millimeters down to 2~$\mu$m.

At liquid helium temperatures, samples with $L\leq$100$~\mu$m exhibit
reproducible conductance fluctuation (CF) as function of the chemical
potential. These are observable in the conductance versus gate-voltage
$G(V_{g})$ scans as a fluctuating component of $G,$ defining a sample-specific
pattern that changes form when the sample is thermally re-cycled. The CF
pattern appears whether or not the underlying glassy system was allowed to
equilibrate prior to taking a $G(V_{g})$ sweep. A $G(V_{g})$ scan taken after
the sample equilibrates reveals the usual memory-cusp, which is the main
earmark of the electron-glass \cite{1,4,5,6} while the CF appear just as a
superimposed structure with no sign of inter-modulation effects. The two types
of conductance modulation observable in the $G(V_{g})$ scans, namely the CF
and memory-cusp, are shown to be distinct physical phenomena; they are
associated with different spatial scales, and quite different time scales.

\section{Experimental}

\subsection{Sample preparation and measurement techniques}

The $In_{2}O_{3-x}$ films used in this work were e-gun evaporated on a
$\operatorname{Si}O_{2}$ insulating layer ($0.5~\mu$m thick) thermally grown
on a
$<$%
100%
$>$
Si wafer. The choice of $In_{2}O_{3-x},$ the crystalline version of
indium-oxide for this study was motivated by its essentially constant
stoichiometry (due to crystal chemistry constraints) hence a fairly constant
carrier concentration. As was shown previously using the amorphous version, a
change in carrier concentration may lead to a large variance in the glassy
features \cite{6}, which might mask the size dependence. The Si wafer was
boron doped and had resistivity $2\cdot10^{-3}~\Omega$cm$,$ deep in the
degenerate regime. It thus could be used as an equipotential gate electrode
for a low-temperature measurement where the $In_{2}O_{3-x}$ served as the
active layer in a 3-terminal device. Lateral dimensions of the samples were
controlled using a stainless steel mask (for samples larger than $0.5~$mm), or
optical lithography (for sizes in the range $30-200~\mu$m), and e-beam
lithography for samples smaller than $30~\mu$m. Samples used in this study had
length $L$ and width $W$ that ranged from $2~\mu$m to $2~$mm$,$ and typical
thickness $d=55\pm5~$\AA . The `source' and `drain' contacts to the
$In_{2}O_{3-x}$ film were made from thermally evaporated $\approx500~$\AA
$~$thick gold films. Fuller details of sample preparation and characterization
are given elsewhere \cite{5}.

Conductance measurements were performed using two terminal ac technique,
employing ITHACO-1211 current preamplifier and PAR-124A lock-in amplifier. In
the MOSFET-like samples, gate voltage sweeps were affected by charging a
$10~\mu$F capacitor with a constant current source. All the measurements
reported here were performed with the samples immersed in liquid helium at
$T=4.11~$K held by a 100 liters storage-dewar. This allowed long term
measurements without disturbing the samples as well as a convenient way to
maintain a stable bath temperature. These requirements are of particular
importance for studies of glassy systems where sample history may influence
time dependent measurements, as was demonstrated in previous studies
\cite{10}. Care was taken to use small bias in the conductance measurements to
ensure linear-response conditions, and in general that was the case. The only
exception is in the 2~$\mu$m samples where some deviations from Ohms law were
observable even at the lowest bias we managed to measure with a reasonable
signal to noise ratio (see also \cite{11}).

\section{Results and discussion}

\subsubsection{Conductance fluctuations}

The emphasis in this work is the evolution of the glassy features as the size
of the system is reduced. These features are observable by performing
conductance measurements employing low frequency techniques, and therefore
flicker noise becomes a major problem in reducing the system size. Mesoscopic
samples naturally exhibit a larger 1/f-noise than macroscopic specimen with
similar resistance (and measured at the same temperature). It was shown
elsewhere that the 1/f-noise magnitude in these samples scales with the
inverse sample volume as expected of independent fluctuators \cite{11}. It
turns out that for $L$ smaller than few microns the noise overshadows some of
the glassy features, which made it difficult to study samples with $L,W<2~\mu
$m$.$ The other feature that becomes progressively more prominent as $L$ goes
smaller is a set sample-specific pattern that modulates the conductance. These
`conductance fluctuations' (CF) are seen in the gate-voltage sweeps $G(V_{g}%
)$, which is the main method to study the various glassy properties. When the
sample size is reduced, the CF, much like the 1/f-noise, eventually masks the
glassy effects. In this subsection we describe the phenomenology of the CF and
their relation to the glassy dynamics on one hand and to the 1/f noise on the
other hand.

Figure 1 shows $G(V_{g})$ scans for two of the studied samples, illustrating
the reproducibility of the CF pattern upon repeated scans (figure 1a and
figure 1b). There are obviously some deviations from perfect reproducibility,
but they are always within the level expected from the observed magnitude of
the 1/f noise. (Namely, the deviation, judged by the cross-correlation between
a traces taken at time $t_{1}$ and $t_{2},$ never exceeded the magnitude of
the noise estimated at $[t_{2}-t_{1}]^{-1}).$ Another general trend is that
the relative reproducibility appears to be better the smaller is the sample
area. The degree of reproducibility was not sensitive to the scan rate of
$V_{g}$ (tested over the range of 0.1-100~mV/s) so we are inclined to believe
that irreproducibility is due to an inherent process, probably related to the 1/f-noise.

The relative magnitude of both, the CF and the 1/f-noise depends on the sample
resistance and temperature, as well as on the sample size. The dependence of
the rms $\Delta G$ associated with the CF on the sample area $A=L\cdot W$ for
samples with similar resistances and measured at the same temperature is shown
in figure 1c (compare with fig.2 in reference \cite{11}).

The two traces in figure 1a were taken one after the other starting
\textit{immediately} after quench-cooling the sample from $T>50~$K to
$T=4.1~$K$.$ Figure 2, on the other hand, shows $G(V_{g})$ traces for a sample
taken \textit{after} it was allowed to equilibrate for 12 hours with $V_{g}$
held at $V_{g}^{0}=0~$V. For the second and third quench-cool runs included in
the figure, the sample was first warmed up to $T\approx70~$K, held there for
$10\sec,$ and then quench-cooled to $4~$K and allowed to relax again for 12
hours prior to taking a $G(V_{g})$ sweep. In all these instances the
$G(V_{g})$ curves show a `memory-cusp' centered at the gate voltage at which
the sample equilibrated, in addition to a distinctive set of CF (figure 2a).
Note that, for each quench, a \textit{quench-specific} set of CF is obtained
while the average value of $G$ is nearly identical (figure 2a), and so is the
shape of the associated memory-cusp (figure 2b).

Before discussing the interplay between the CF and the memory-cusp, the origin
of the CF in these samples needs to be clarified. There is some similarity in
behavior between the CF observed in our $G(V_{g})$ and the UCF phenomenon
familiar from the weak localization (diffusive) regime \cite{12}. Most notably
in that the fluctuation pattern produced is like a set of `finger-prints'
characteristic of a particular structural, frozen-in configuration. However,
the underlying mechanisms for the two phenomena are different; The UCF is a
quantum interference effect usually produced by sweeping a magnetic field $H$
that modulates $G$ via the Aharonov-Bohm effect (similar behavior has been
observed in gate experiments on diffusive samples, likewise related to an
interference effect \cite{13}). The CF observed here is associated with a
mechanism offered by Lee \cite{14} to account for experiments on quasi-one
dimensional hopping samples \cite{15}. In terms of variable range hopping our
samples are therefore effectively two-dimensional (A typical hopping length in
our samples at the temperatures used in this work is $\approx200~$%
\AA ~\cite{16}, which is larger than the thickness $\simeq55~$\AA ). The
underlying physics of Lee's mechanism however, does not depend on the
particular dimensionality: The current in hopping systems is carried in a
percolation-network and controlled by a relatively small number of `critical'
or `bottleneck' resistors \cite{17,18,19}. Like any other resistor in the
system, the values of these resistors are given by the Miller-Abrahams
expressions \cite{20} $R_{ij}\propto\exp[\frac{2r_{ij}}{\xi}+\frac
{|E_{i}-E_{j}|+|E_{i}-\mu|+|E_{j}-\mu|}{k_{B}T}]$ where $r_{ij}$ is the
inter-site distance in space, $\xi~$is the localization length, $~E_{i}$ ,
$E_{j}$ are the site energies, and $\mu$ is the chemical potential. Therefore
the actual resistance of these circuit elements depend (among other things) on
the value of the chemical potential. Sweeping the gate voltage causes the
chemical potential to change thereby re-shuffling the values of the resistors
in the system. The CF pattern in this picture reflects the process by which
some critical resistors in the percolation network are replaced by other
critical resistors as the chemical potential is varied \cite{14}. The fact
that by going back and forth with $V_{g}$ the pattern reproduces itself
suggests that there is an underlying backbone structure of sites that are
fairly stable, most likely the positions of the ions that control the disorder
in the system. This is the analog of the distribution of scattering centers
that determine the magnetic fingerprints in the UCF effect. In both phenomena
the underlying structure changes upon thermal recycling, and in both the
deviations from perfect reproducibility are associated with non-stationary
potential disorder which in turn is believed to be the main cause of the 1/f
noise \cite{21}.

The relative magnitude of the CF for a given 2D sample is determined by two
factors. The first is related to the basic conductance swing $\Delta G$
associated with a fluctuation in the hopping regime, which is of the order of
the system conductance $G.$ This is a characteristic feature of the strongly
localized regime, (which in the present case of 2D samples means sheet
resistance $R_{\square}$ that fulfills~$R_{\square}\gg\frac{\hbar}{e^{2}}$).
In other words, for a system with size $L$ smaller or equal to the scale
relevant for the phenomenon observed, $\Delta G/G$ is of the order unity
\cite{22}. This should be contrasted with the `universal' conductance
fluctuation $\Delta G\approx e^{2}/h$ ~characterizing the respective situation
in weak localization (which holds for $G\gg e^{2}/h$ and thus $\Delta G/G\ll1$
being the typical case$).$ This, incidentally is the main reason why CF
(whether due to an interference effect or an energy-shift mechanism), and
conductance noise are much more prominent in the hopping regime than they are
in the diffusive one. The second factor that controls the fluctuation
amplitude $\Delta G/G$ is how many independent fluctuators the
current-carrying network contains. This is essentially determined by the
square root of the number of critical resistors in the sample $N\approx
\frac{L\cdot W}{\mathcal{L}^{2}}=\frac{A}{\mathcal{L}^{2}}$ where
$\mathcal{L}$ is the percolation radius \cite{17,18,19} and thus,
$\frac{\Delta G}{G}\approx\mathcal{L}A^{-1/2}$. This law is depicted in figure
1c and yields $\mathcal{L}\approx$ 0.3$~\mu$m (as can be easily seen from the
figure by the extrapolated value of the best fit line to $\frac{\Delta G}%
{G}\approx1)~$ This value for the percolation radius is quite reasonable for
samples with $R_{\square}$ in the range used (and for $T\approx4~$K). Not
surprisingly, a similar dependence on $\mathcal{L}$ was observed for the 1/f
noise magnitude \cite{11}.

It should be noted that $\mathcal{L}$ is disorder-dependent quantity
\cite{17,18,19} and for a fixed temperature, as is the case for the data
presented here, the relative amplitude of the CF (as well as the 1/f-noise)
increases with disorder. The relative amplitude of the memory cusp also
increases with disorder \cite{23} making it hard to separate the
electron-glass from the CF phenomenon by this parameter. The features that
clearly distinguish between these phenomena are the dependence on the system
size, and their different temporal behavior as demonstrated below.

The glassy effects seem to be simply superimposed on the CF and noise without
any noticeable intermodulation between these phenomena. For example, figure 3
demonstrates that the reproducibility of the CF is unaffected by the presence
of the memory-cusp. In figure 3a we compare the $G(V_{g})$ taken immediately
after cool-down with the one after the sample was allowed to relax under
$V_{g}=6~$V$,~$which now includes a memory-cusp superimposed on the CF
pattern. Indeed, removing a `simulated' cusp (figure 3b) from the `relaxed'
$G(V_{g})$, the two curves show the same degree of CF registry in the voltage
region where the original cusp appeared as in the regions outside it (figure 3c).

While the phenomenology associated with the CF appears to have several common
features with the 1/f noise, the electron-glass effects seem to be controlled
by a different spatial scale, and quite a different time scale. Neither the
width nor the relative amplitude of the memory-cusp change when $L$ is
decreased down to the smallest scale used in our experiments. This is in
contrast with the behavior of the CF and the 1/f noise, which is also why the
glassy effects become difficult to resolve below a certain scale. The huge
disparity in the dynamics between the CF and the electron-glass is illustrated
by the experiment described in figure 4. Starting from a relaxed state at the
cool-down $V_{g}=V_{g}^{o}$, multiple sweeps of the gate voltage over an
interval straddling $V_{g}^{o}$ yield $G(V_{g})$ traces that clearly show that
the cusp amplitude decays with time \textit{much} faster than the time it
takes the CF pattern to change appreciably. In fact, note that for the last
3-4 sweeps, the amplitude of the memory-cusp has already decayed to a level
comparable with the amplitude of some of the wiggles associated with the CF,
which appear to be fairly stationary throughout the entire experiment.

\subsubsection{Memory and dynamics}

In addition to the memory cusp discussed above, essentially all the other
characteristic features of the electron glass are observable in the samples
that approach the mesoscopic regime. Figure 5 compares the aging behavior of a
8 by 18$~\mu$m sample with that of a 1x1~mm sample both using the
`gate-protocol' \cite{5,10} with similar parameters and similar waiting times.
The same aging protocol was used involving the following procedure. The sample
is allowed to relax for 12 hours at $T=4~$K under a gate voltage $V_{g}^{o}$
reaching a near-equilibrium conductance $G_{0}$ (which is used as baseline for
the aging protocol). Then the gate voltage is quickly changed (typically over
5 seconds) from $V_{g}^{o}$ to $V_{g}^{n}~$and is held there for $t_{w}$ after
which $V_{g}$ is re-instated at $V_{g}^{o}$. Figure 5 is a plot of the excess
conductance (normalized to the equilibrium conductance $G_{0})$ measured after
$V_{g}$ is set back to $V_{g}^{o}.$ The figure shows such $\Delta G/G$ curves
as function of time for different $t_{w}$ (figure 5a and 5c). The same data
are plotted as function of $t/t_{w}$ to illustrate simple-aging behavior
(figure 5b and 5d), which is almost identical for the `mesoscopic' and
macroscopic sample. The only apparent difference is a more pronounced noise
superimposed on the data of the small sample.

Down to sample sizes of 8~$\mu$m by 18~$\mu$m, and for the $R_{\square}$
values used in this study (typically, 5-200\ M$\Omega$ at 4\ K), the amplitude
of the 1/f-noise and the CF were small enough making it possible to observe a
memory-cusp and to perform a two-dip-experiment, which is one of the better
defined techniques to get a measure of the glassy dynamics \cite{6}. The
results of such an experiment is shown in figure 6. Note that the
characteristic relaxation time $\tau,$ as defined by the two-dip experiment,
for this sample (figure 6b) is of order $10^{3}\sec.$ This is quite similar to
the values typically obtained for samples of $In_{2}O_{3-x}$ of millimeter
size \cite{3}, which means that down to $\approx10~\mu$m the dynamics does not
change from its macroscopic value. Unfortunately, it was not feasible to
perform a two-dip experiment on smaller samples because the magnitude of the
CF became larger than the modulation associated with the memory-cusp.

To be able to make a systematic study of the dependence of glassy dynamics on
sample size including samples smaller than $10~\mu$m we resorted to the
'single-conductance-excitation' method. This is based, as a first step, on
monitoring the relaxation of the excess conductance created e.g., by a $V_{g}$
switch from $V_{g}^{o}$ to $V_{g}^{n}$. The excess conductance follows the
logarithmic law \cite{5,10}:%

\begin{equation}
\Delta G(t)=\Delta G(t_{0})-a\log(t/t_{0}) \label{1}%
\end{equation}

(where $\Delta G(t_{0})$ is the initial amplitude of the excess conductance
attained, just after $V_{g}$ is switched to $V_{g}^{n}$ (within the experiment
resolution-time $t_{0}$, typically $1\sec$).\texttt{ }Conductance relaxation
curves exhibiting the logarithmic law for sample sizes down to 8x18~$\mu$m are
illustrated in figure 7.The systematic increase of noise as the samples
dimension is reduced is clearly reflected from this figure yet the $\log(t)$
law is reasonably well defined in these samples and may be used to get a
measure of dynamics as explained below. For the 2x2$~\mu$m samples however,
the 1/f-noise amplitude is often larger than $\Delta G(t_{0})$ and the
relaxation process may be obscured to the degree that even the sign of
$\partial\lbrack\Delta G(t)]/\partial t$ is in doubt.The noise for a typical
2x2$~\mu$m sample is shown in figure 8, and in the time trace one can see
time-periods during which the conductance may drift in either direction from
its mean value by 2-4\% which is comparable to the $\Delta G(t_{0})$
associated with (macroscopic) samples with similar parameters \cite{10}. Due
to the $\log(t)$ law, over 10$^{2}$-10$^{3}\sec$ a substantial part of $\Delta
G$ dies out, and therefore even the sign of $\Delta G$ versus $t$ becomes
difficult to ascertain in this situation. Two relaxation curves illustrating
this unwieldy behavior of $\Delta G(t)$ for typical single-runs of small
sample are shown in figure 9a and 9b. Apparently, the only way to get a more
manageable $\Delta G(t)$ in this realm of sample size is to use averaging.
Figure 9c shows the relaxation curve for the same 2x2$~\mu$m sample averaged
over 53 individual runs that were acquired by performing multiple
excitation-relaxation runs. This was realized by using two end-points for
$V_{g}$ along the gate voltage range, and monitor the relaxation of $G$ at
each end-point for a time $t^{\ast}$ after $V_{g}$ was switched from the other
end-point. The two end-points $V_{g}^{1}$ and $V_{g}^{2}$ were chosen to be
far enough from one another such that the voltage swing $|V_{g}^{1}-V_{g}%
^{2}|$ was considerably bigger than the width of the memory-cusp to minimize
history effects \cite{10}. For the same reason the data used in figure 9 are
limited to time-spans smaller than $t^{\ast}.$ The resulting relaxation curve
(figure 9c), except for being still much more noisy, is quite similar to the
relaxation curves of the larger samples in that it follows the logarithmic
law, and with similar dynamics as is shown next.

To extract a measure of the dynamics from the $G(t)$ curves such as in figures
7 and 9, one needs first to subtract the equilibrium value of $G,$ then take
into account the equilibrium field effect $\Delta G_{eq}.$ The latter is the
change in $G$ due to the change in the thermodynamic density of states
associated with the $V_{g}^{o}\rightarrow V_{g}^{n}$ switch \cite{25}. The
value of $\Delta G_{eq}$ is estimated separately for each sample using the
anti-symmetric part of the field effect (see reference \cite{25} for details).
Then, normalizing by $\Delta G(t_{0})$ yields a logarithmic curve for $\Delta
G(t)$ with the boundary condition $\Delta G(t_{0})\equiv1$ from which a the
relaxation rate for $\Delta G$ is obtained as percent change per time-decade.
With the above normalization this value is a proper measure of the system
dynamics in that it allows comparison between different samples and/or
different conditions of measurement. Figure 10 summarizes the results of this
procedure for samples with sizes ranging between 1x1~mm to 2x2$~\mu$m$,$ all
tested at $T=4.1~$K$,$ and having similar resistances.

In principle, a similar averaging stratagem as that described above to deal
with the 1/f noise problem may be employed to facilitate the observation of
the memory cusp in small samples. This would entail averaging over data
obtained in many quench-cooling events, each followed by a long period of
relaxation (to build up the cusp). Such a procedure however is prone to sample
damage in addition to being prohibitively time consuming.

To summarize, we have shown in this work that the dynamics of the
electron-glass has a different time and spatial scale than those associated
with the mesoscopic conductance fluctuations. These conclusions are inferred
from the following experimental results:

\begin{itemize}
\item The CF pattern is fully developed immediately after cooling the sample,
while the memory-cusp amplitude takes many hours to build.

\item Starting from a cold and well-equilibrated sample, continuously sweeping
$V_{g}$ causes the memory-cusp amplitude to diminish with a typical time scale
of hours while the CF visibility remains essentially intact for many days.

\item Thermally cycling the sample creates a different pattern of CF while the
shape and relative magnitude of the memory-cusp is the same (as long as the
average conductance of a given sample is the same).
\end{itemize}

Inasmuch as the CF pattern is taken as a snapshot of the underlying potential
landscape, these findings seem to indicate that the electron-glass dynamics
may be treated as essentially independent of the lattice dynamics although in
terms of energy exchange (and dissipation) the two systems are coupled. This
is not a trivial result as there is no reason to assume that the lattice
potential is stationary over the time scale relevant for the electron glass.
In fact, ions, atoms, and perhaps group of atoms, are presumably mobile even
at cryogenic temperatures, thus producing potential fluctuations \cite{26}.
These are believed to be the main source of the 1/f noise which includes time
scales that overlap those typical of the electron glass dynamics. It is
therefore plausible to expect that the two phenomena affect one another to
some degree. For example, one intuitively expects that the conductance noise
will be modified when the system is relaxing from an excited state (or, more
generally, when it is out of equlibrium). In the present case however, where
$\Delta G$ in the excited state is typically few percents of the equilibrium
conductance, this seems to be a small effect that so far escaped our detection
\cite{27}. An attempt to find a change in the 1/f noise characteristics when
the system is deliberately taken out of equilibrium yielded a null result.
Such an experiment was carried out by keeping the sample in an excited state
for a sufficiently long time to allow adequate averaging of the noise spectrum
\cite{28}. This steady-state, out-of-equilibrium situation was achieved by a
continuous sweep of the gate voltage over a certain voltage interval. Data for
$G(t)$ were taken after the memory-cusp was erased by allowing this process to
proceed for two days. The noise characteristics were compared with those taken
with a fixed-gate situation for which data were taken after the sample was
allowed to relax (and build a memory-cusp). In either case the noise spectrum
was obtained by Fourier transforming 25 conductance time-traces each composed
of 1024 points taken with 1 second resolution. It turns out that, within the
error of the procedure, the 1/f noise magnitude and spectrum were essentially
the same for the `near-equilibrium' and the `excited' sample. Full details of
this experiment as well as the implications to the general issue of flicker
noise in the hopping regime will be given elsewhere \cite{29}.\ 

Finally, throughout the size-range where the memory-cusp could be resolved
over the background of the CF and the 1/f noise, its amplitude and
characteristic width\ do not depend on sample size. Likewise, the dynamics
associated with conductance relaxation is independent of sample size down to
2$~\mu$m. These findings suggest that the relevant length scale for the
electron glass is considerably smaller than the sample sizes used in this study.

The authors are grateful for useful discussion with A. Efros, and M. Pollak.
This research was supported by a grant administered by the US Israel
Binational Science Foundation and by the Israeli Foundation for Sciences and Humanities.

{\Large Figure captions}

1. $G(V_{g})$ traces for two of the samples with different sizes. In each case
two traces were taken one after the other starting immediately (within
$\simeq20\sec.)$ after quench-cooling the sample from $T\approx50~$K to the
measurement temperature of 4.1~K. Scan rate was 10~mV/s. Samples sheet
resistance $R_{\square}=57.5~$M$\Omega$ and 4$~$M$\Omega$ for (\textbf{a}) and
(\textbf{b}) respectively. Graph (\textbf{c}) shows the dependence of the rms
conductance fluctuations on the sample area. This is based on samples with
similar thickness, and $R_{\square}$ (in the range of 3-10$~$M$\Omega$), each
point is the average value of 2-3 samples. The dashed line is a best fit to
$\Delta G/G\propto A^{-1/2}$ law.

2. (\textbf{a}) $G(V_{g})$ traces for three consecutive quench-cooling runs of
the same sample ($R_{\square}=57.5$M$\Omega,$ $L=25~\mu$m, $W=100~\mu$m). For
each trace the sample was heated to $\approx70~$K for 10 seconds, then cooled
to 4$~$K and allowed to equilibrate for $\approx$12 hours holding $V_{g}=0~$V
prior to taking a scan (scan rate 10~mV/s.). Graph (\textbf{b}) is the same
data normalized for each run to $G(0).$ Note that the memory-cusp is nearly
identical in all three runs while the CF pattern is different. Also, note that
the value of the conductance (that is presumed to reflect the average
disorder), is nearly the same in all three quenches (the difference in $G_{0}$
between the 1st and 2nd quench is negligible with respect to the variation of
$G$ affected by cooling from 70\ K to 4\ K).

3. (\textbf{a}) The fluctuating part of several $G(V_{g})$ traces (i.e., after
subtracting the equilibrium background) for a sample with $R_{\square}%
=12.1~$M$\Omega,$ $L=25~\mu$m, $W=20~\mu$m. The first trace, measured
immediately (within $\simeq20\sec.)$ after quench-cooling the sample from
$T\approx70~$K to the measurement temperature of 4.1$~$K (empty circles). A
second trace (full triangles) taken after the sample was allowed to
equilibrate under $V_{g}=6~$V for $\approx$12 hours. The data for the latter
scan are shown in (\textbf{b}) along with a fitted (to a Lorentzian)
memory-cusp (dashed line) that when subtracted from these data results in the
full-triangle trace compared in (\textbf{c}) with the first scan. All traces
taken with rate of 10~mV/s.

4. $G(V_{g})$ traces taken consecutively starting from $V_{g}=0~$V where the
sample equilibrated for $\simeq$12 hours, then swept continuously in the
interval $V_{g}=-3$V to $V_{g}=+3$V with a scan rate of 10$~m$V/s. The traces
are labeled by their "age" relative to the starting time of the experiment.
$\frac{\Delta G}{G}$ is $\frac{[G(V_{g})-G_{0}(0)]}{G_{0}(0)}$ where $G_{0}$
is the equilibrium conductance. Sample parameters: $R_{\square}=37.5~$%
M$\Omega,$ $L=80~\mu$m, $W=200~\mu$m.

5. Aging experiment using the gate protocol comparing the results on 1x1$~$mm
sample (\textbf{a} and \textbf{b}, $R_{\square}=57~$M$\Omega),$ with those for
$L=$8$~\mu$m and $W=18~\mu$m (\textbf{c} and \textbf{d}, $R_{\square}%
=3.5~$M$\Omega)$. See text for the protocol details.

6. Two-dip experiment with a $L=8~\mu$m, $W=18~\mu$m sample ($R_{\square}%
=55~$M$\Omega).$ Top graph shows $G(V_{g})$ traces taken at different times;
starting from equilibrated position under $V_{g}^{o}=0~$V$,$ an initial scan
is taken showing a well developed memory-cusp at this $V_{g}.$ The gate
voltage is then swept to and parked at $V_{g}^{n}=1.85~$V for the rest of the
experiment and $G(V_{g})$ traces taken at later times as indicated (traces are
displaced along the ordinate for clarity). Note the appearance of a new
memory-cusp at $V_{g}^{n}$ becoming more prominent with time while the
visibility of the old cusp at the original equilibration voltage $V_{g}^{o}$
decreases with time$.$Scan rate for all traces was 10~mV/s. Bottom graph shows
the decay of amplitude for the old cusp (circles) and the build-up of
memory-cusp amplitude at the new gate voltage position (triangles). Amplitudes
are measured relative to the `natural' background of $G(V_{g}).$

7. Relaxation of the sample conductance following an excitation by switching
the gate voltage from a $V_{g}^{o}$ at which the sample was equilibrated to
$V_{g}^{n}$ (the typical swing of $V_{g}~$used in these experiments
$|V_{g}^{o}-V_{g}^{n}|~$was in the range $2-10~$V). Data are shown for samples
with different lateral dimensions as indicated and with $R_{\square}$ of
35.5$~$M$\Omega,$ 69.2$~$M$\Omega,$ and 175$~$M$\Omega$ for the sample in
(\textbf{a}), (\textbf{b}), and (\textbf{c}) respectively.

8. Conductance versus time (at equilibrium) for a sample with 2x2$~\mu$m
lateral dimensions and with $R_{\square}$ of 56$~$M$\Omega.$Note the
accentuated noise level. The inset shows the power spectrum averaged over 14
time traces each 1024 $\sec.$ long.The dashed line is a best fit to
1/f$^{\alpha}$ law ($\alpha\approx1.15$).

9. Relaxation of the sample conductance following an excitation by switching
the gate voltage from a $V_{g}^{o}=-7~$V at which the sample was equilibrated
to $V_{g}^{n}=+7~$V for the same sample as in figure 8 (\textbf{a}) and
(\textbf{b}) are two typical single-runs while (c) is the result of averaging
over 53 such traces that were obtained by alternate switching between
$V_{g}^{o}$ and $V_{g}^{n}$ (see text for details).

10. The relaxation rate of the excess conductance produced as in figure 7 and
8 for samples with various sizes and with similar values of $R_{\square},$ all
measured at $T=4.1~$K$.$


\begin{thebibliography}{99}                                                                                               %


\bibitem {1}M. Ben Chorin, Z. Ovadyahu and M. Pollak, Phys. Rev. B\textbf{
48}, 15025 (1993).

\bibitem {2}G. Martinez-Arizala, D. E. Grupp, C. Christiansen, A. Mack, N.
Markovic, Y. Seguchi, and A. M. Goldman, Phys. Rev. Lett., \textbf{78}, 1130
(1997). G. Martinez-Arizala, C. Christiansen, D. E. Grupp, N. Markovic, A.
Mack, and A. M. Goldman, Phys. Rev. B\textbf{ 57}, R670 (1998).

\bibitem {3}Z. Ovadyahu and M. Pollak, Phys. Rev. Lett., \textbf{79}, 459 (1997).

\bibitem {4}T. Grenet, Eur. Phys. J, \textbf{32}, 275 (2003).

\bibitem {5}A. Vaknin, Z. Ovadyahu, and M. Pollak, Phys. Rev. B\textbf{ 65},
134208 (2002).

\bibitem {6}A. Vaknin, Z. Ovadyhau, and M. Pollak, Phys. Rev. Lett.,
\textbf{81}, 669 (1998).

\bibitem {7}M. M\"{u}ller and L. B. Ioffe, Phys. Rev. Lett. \textbf{93},
256403 (2004).

\bibitem {8}Sh. Kogan, Phys. Rev. B \textbf{57}, 9736 (1998).

\bibitem {9}T. Kawasaki, and S. Miyashita, Journal of Magnetism \& Magnetic
Mater., \textbf{104}, 1595 (1992); A. Bunde, S. Havlin, J. Klafter, G. Graff,
A. Shehter, Phys. Rev. Lett., \textbf{78}, 3338 (1997); Andrea Montanari, and
Guilhem Semerjian, J. Stat. Phys. \textbf{125}, 23 (2006).

\bibitem {10}Z. Ovadyahu, and M. Pollak, Phys. Rev. B\textbf{ 68}, 184204 (2003).

\bibitem {11}V. Orlyanchik, V. I. Kozub, and Z. Ovadyahu, Phys. Rev. B
\textbf{74}, 235206 (2006).

\bibitem {12}S. Washburn and R. A. Webb, Reports on Prog. in Phys.,
\textbf{55}, 1311 (1992), and references therein.

\bibitem {13}W. J. Skocpol, P. M. Mankiewich, R. E. Howard, L. D. Jackel, and
D. M. Tennant, Phys. Rev. Lett. \textbf{56}, 2865 (1986).

\bibitem {14}P. A. Lee, Phys. Rev. Letters, \textbf{53}, 2042 (1984).

\bibitem {15}A. B. Fowler, A. Hartstein, and R. A. Webb, Phys. Rev. Lett.,
\textbf{48}, 196 (1982); W. J. Skocpol, L. D. Jackel, E. L. Hu, R. E. Howard,
and L. A. Fetter, Phys. Rev. Lett. \textbf{49}, 951(1982); R. F. Kwasnick, M.
A. Kastner, J. Melngailis, and P. A. Lee, Phys. Rev. Lett. \textbf{52}, 224 (1984).

\bibitem {16}O. Faran and Z. Ovadyahu, Phys. Rev. B\textbf{ 38}, 5457 (1988);
O. Faran and Z. Ovadyahu, Solid State Comm., \textbf{67}, 823 (1988); D.
Shahar and Z. Ovadyahu, Phys. Rev. Lett. \textbf{64}, 2293 (1990).

\bibitem {17}B. I. Shklovskii, A.L. Efros, Sov. Phys. JETP \textbf{33}, 468 (1971).

\bibitem {18}M. Pollak, J. Non-Cryst. Solids \textbf{11}, 1 (1972).

\bibitem {19}V. Ambegaokar, B. I. Halperin, J. S. Langer, Phys. Rev.
\textbf{B4}, 2612 (1971)

\bibitem {20}A. Miller and E. Abrahams, Phys. Rev. \textbf{120}, 745 (1960).

\bibitem {21}P. A. Lee and A. D. Stone, Phys. Rev. Lett. \textbf{55}, 1622
(1985); S. Feng, P. A. Lee, and A. D. Stone, Phys. Rev. Lett. \textbf{56},
1960 (1986).

\bibitem {22}Y. Imry, Europhys. Lett. \textbf{1}, 249 (1986); F. P. Milliken
and Z. Ovadyahu, Phys. Rev. Lett., \textbf{65}, 911 (1990); Z. Ovadyahu, Waves
in Random Media, \textbf{9}, 241 (1999) and references therein.

\bibitem {23}M. Pollak and Z. Ovadyahu, J. de Physique I France, \textbf{7},
1595 (1997).

\bibitem {24}A. Vaknin, Z. Ovadyhau, and M. Pollak, Phys. Rev. Lett.,
\textbf{84}, 3402 (2000).

\bibitem {25}Z. Ovadyahu, Phys. Rev. B \textbf{73}, 214208 (2006).

\bibitem {26}Movements of these objects, commonly referred to as
two-state-systems, may be triggered by electronic transitions such as changes
in bonding angle. Therefore these are as much `electronic' as they are
`atomic'. The only distinction we make is based on whether the objects
actually contribute to macroscopic dc current or just modulate it. The
possible role of ions movement and its relation to the electron glass dynamics
was discussed before vis-a-vis the dependence of the dynamics on disorder and
magnetic field \cite{3}.

\bibitem {27}Note that re-normalization of hopping rates due to the coupling
of electrons to two-state-systems may still be at work and that may affect
e.g., the relaxation rate of $\Delta G$. This issue is now under study.

\bibitem {28}The relaxation of $\Delta G$ in a `one-shot' experiment (as for
example in figure 7) leaves very little time for a measurement of 1/f-noise
while the system is in the `excited' state; due to the logarithmic law most of
the excess conductance (which is small anyhow relative to the equilibrium
value) is gone by $\Delta t\approx100-500\sec.$

\bibitem {29}Z. Ovadyahu (unpublished).
\end{thebibliography}
\end{document}